\documentclass[aps,prl,twocolumn,superscriptaddress,showpacs]{revtex4}
\usepackage{bm}
\usepackage{graphicx}
\usepackage{amssymb,amsmath,amsbsy,amsgen,amsfonts}     
\usepackage{dcolumn}
\usepackage{amsthm}
\usepackage{mathrsfs}
\usepackage{latexsym}
\usepackage{array}
\usepackage{amstext}
\usepackage{txfonts}

\usepackage{epstopdf} 

\newcommand{\ncd}{\newcommand}
\ncd{\QC}{$\mbox{QC}_{\cal{C}}$}
\ncd{\QCpr}{${\mbox{QC}_{\cal{C}}}^\prime\;$}
\ncd{\QCns}{$\mbox{QC}_{\cal{C}}$}
\ncd{\QCprns}{${\mbox{QC}_{\cal{C}}}^\prime$}
\ncd{\cskN}{{|\phi_{\{\kappa\} } \rangle}_{{\cal{C}}_N}}
\ncd{\cskNpr}{{|\phi_{\{\kappa^\prime\} } \rangle}_{{\cal{C}}_N}}
\ncd{\cskNtil}{{|\phi_{\{\tilde{\kappa} \} }
\rangle}_{{\cal{C}}_N}}
\ncd{\csk}{{|\phi_{\{\kappa\} }
\rangle}_{\cal{C}}}
\ncd{\csktil}{{|\phi_{\{\tilde{\kappa} \} }
\rangle}_{\cal{C}}}
\ncd{\cskf}{|\phi_{\{\kappa\} }
\rangle_{\cal{C}}}
\ncd{\csktilf}{|\phi_{\{\tilde{\kappa} \} }
\rangle_{\cal{C}}}
\ncd{\bracsk}{\mbox{}_{\cal{C}}\langle\phi_{\{\kappa\} }|}
\ncd{\bracsktil}{\mbox{}_{\cal{C}}\langle\phi_{\{\tilde{\kappa} \}
}|} \ncd{\nbracsk}{\mbox{}_{\cal{C}}\langle\phi_{\{\kappa\} }}
\ncd{\nbracsktil}{\mbox{}_{\cal{C}}\langle\phi_{\{\tilde{\kappa}
\} }} \ncd{\cs}{|\phi \rangle_{\cal{C}}\;} \ncd{\csns}{|\phi
\rangle_{\cal{C}}}
\ncd{\nbgh}{\text{nghb}} 
\ncd{\Sab}{S^{ab}}
\ncd{\Sba}{S^{ba}}
\ncd{\ds}{\displaystyle} \ncd{\ovl}{\overline}

\ncd{\SABC}{S^{ABC}}
\ncd{\Sbc}{S^{bc}}

\newcommand{\ket}[1]{\left\vert{#1}\right\rangle}
\newcommand{\ketbra}[2]{|#1\rangle \langle#2|}

\newcommand{\be}{\begin{equation}}
\newcommand{\ee}{\end{equation}}
\newcommand{\ba}{\begin{array}}
\newcommand{\ea}{\end{array}}
\newcommand{\bqa}{\begin{eqnarray}}
\newcommand{\eqa}{\end{eqnarray}}

\begin{document}

\title{Experimental Realization of a One-way Quantum Computer Algorithm Solving Simon's Problem}

\author{M. S. Tame} 
\email{markstame@gmail.com}
\address{University of KwaZulu-Natal, School of Chemistry and Physics, 4001 Durban, South Africa}
\address{National Institute for Theoretical Physics, University of KwaZulu-Natal,Durban 4001,
South Africa}

\author{B. A. Bell}
\affiliation{Photonics Group, Department of Electrical and Electronic Engineering, University of Bristol, Merchant Venturers Building, Woodland Road, Bristol, BS8 1UB, UK}

\author{C. Di Franco} 
\affiliation{Quantum Optics and Laser Science Group, Imperial College London, Blackett Laboratory, SW7 2AZ London, UK}

\author{W. J. Wadsworth}
\affiliation{CPPM, Department of Physics, University of Bath, Claverton Down, Bath, BA2 7AY, UK}

\author{J. G. Rarity}
\affiliation{Photonics Group, Department of Electrical and Electronic Engineering, University of Bristol, Merchant Venturers Building, Woodland Road, Bristol, BS8 1UB, UK}

\date{\today}

\begin{abstract}
We report an experimental demonstration of a one-way implementation of a quantum algorithm solving Simon's Problem - a black box period-finding problem which has an exponential gap between the classical and quantum runtime. Using an all-optical setup and modifying the bases of single-qubit measurements on a five-qubit cluster state, 
key representative functions of the logical two-qubit version's black box can be queried and solved. To the best of our knowledge, this work represents the first experimental realization of the quantum algorithm solving Simon's Problem. The experimental results are in excellent agreement with the theoretical model, demonstrating the successful performance of the algorithm. With a view to scaling up to larger numbers of qubits, we analyze the resource requirements for an $n$-qubit version. This work helps highlight how one-way quantum computing provides a practical route to experimentally investigating the quantum-classical gap in the query complexity model. 
\end{abstract}

\pacs{03.67.-a, 03.67.Mn, 42.50.Dv, 03.67.Lx}

\maketitle

Quantum information science promises to radically change the way we communicate and process information in future devices based on quantum technology~\cite{Gisin,nielsenchuang,Ladd}. One of its major goals is to realize multi-qubit quantum algorithms, involving large numbers of logic gates, that outperform their classical analogues~\cite{algor,Simon}. While there has been steady experimental progress made during recent years in demonstrating basic quantum logic gates in various settings~\cite{Ladd}, the process of piecing them together in order to perform useful algorithms is still far from practical. Demonstrations of few-qubit quantum algorithms therefore play an important role in stimulating further advances in experimental quantum computing (QC) and help open up viable routes toward full-scale quantum information processing. Photonic systems in particular provide a reliable and rapid test bed for emerging quantum technologies with excellent prospects for scalability~\cite{Brien}.

In this work we report the first experimental demonstration of a one-way based implementation of the quantum algorithm solving Simon's Problem (SP)~\cite{Simon}. This is a period-finding problem of great importance in quantum algorithm design as it provides a clear exponential gap between the classical and quantum runtime. It was a motivation for Shor's factoring algorithm~\cite{algor} and has played a major role in the evolution of quantum algorithm design~\cite{Childs}. Here, we exploit the one-way model to experimentally demonstrate SP using a multipartite entangled state, the cluster state, as a resource for running a program represented by single-qubit measurements~\cite{oneway,oneway2,Bristolow}. This measurement-based approach is appealing as it reduces the amount of control one needs over a quantum system to the ability of carrying out measurements only, an important advantage for a number of physical settings, most notably those using ion-traps~\cite{ion1,ion2} and photons~\cite{onewaye,Tame1,DJexp1,DJexp2,Pan}. The one-way model thus continues to generate much interest, both at a theoretical~\cite{onewayt,develop} and experimental~\cite{ion2,onewaye,Tame1,DJexp1,DJexp2,Pan} level. The algorithm we experimentally demonstrate for SP illustrates the unique role that parallelism in QC plays in the speed-up given by quantum algorithms solving classical decision problems. Despite being one of the first quantum algorithms introduced and the first to show that an exponential gap can exist in the runtime between solving problems classically and quantum mechanically, the quantum algorithm for SP has surprisingly never been experimentally demonstrated before. One of the main reasons behind this may be due to the complexity of the quantum circuitry required, even for the smallest instance of the algorithm~\cite{Simon}. Here we show that a measurement-based approach, due to its great flexibilty, finally enables the realization of the algorithm using current technology. Thus, to the best of our knowledge, our work not only represents the first implementation of the algorithm in the promising context of one-way QC, but also the algorithm's first experimental realization in any physical system. 

In our experiment we show that five qubits in a specific cluster state configuration are sufficient to realize key representative functions of the problem's black box acting on a logical four-qubit register; two query and two ancilla qubits. A complex modification to the experimental setup for each function is not necessary, only a small change to the program of measurements, an important advantage over other QC techniques. Our experimental results are in excellent agreement with the theoretical model, demonstrating successful performance of the algorithm in a photonic setting. As photonic technology is a highly promising platform for realizing quantum computing, our demonstration is of great significance for helping open up a practical route to probing larger and more complex quantum algorithms. Along these lines, we also discuss extending our scheme to implement all black box functions for the two-qubit version, not just representatives, in addition to arbitrary sized registers and the resources required. Thus, we show that one-way quantum computing provides a flexible and practical route to experimentally investigating the quantum-classical interface in the query complexity model. 

{\it Model.-} The problem considers an oracle that implements a function mapping an $n$-bit string to an $m$-bit string $f:~\{ 0,1\}^{n} \to \{ 0,1\}^{m}$, with $m \ge n$, where it is promised that $f$ is a $1-1$ type function (each input gives a different output) or $2-1$ type function (two inputs give the same output) with non-zero period $s \in \{ 0,1\}^{n}$ such that for all $x \neq x'$ we have $f(x)=f(x')$ if and only if $x'=x \oplus s$, where $\oplus$ corresponds to addition modulo 2. The problem is to determine the type of the function $f$ and if it is $2-1$, to determine the period $s$. Using classical query methods the probability of solving the problem is given by $p_{s} \leq 1/2 + \delta$, with $\delta=2^{-n/2}$ if one queries the oracle at least $2^{n/4}$ times. As $n \to \infty$, $\delta \to 0$ and the number of queries needed to obtain $p_{s}>1/2$ grows exponentially. Quantum mechanically, the number of queries needed is $O(n)$ to solve the problem with $p_{s}=1$~\cite{Simon}. 
\begin{figure}[t*]
\centerline{\includegraphics[width=8.4cm]{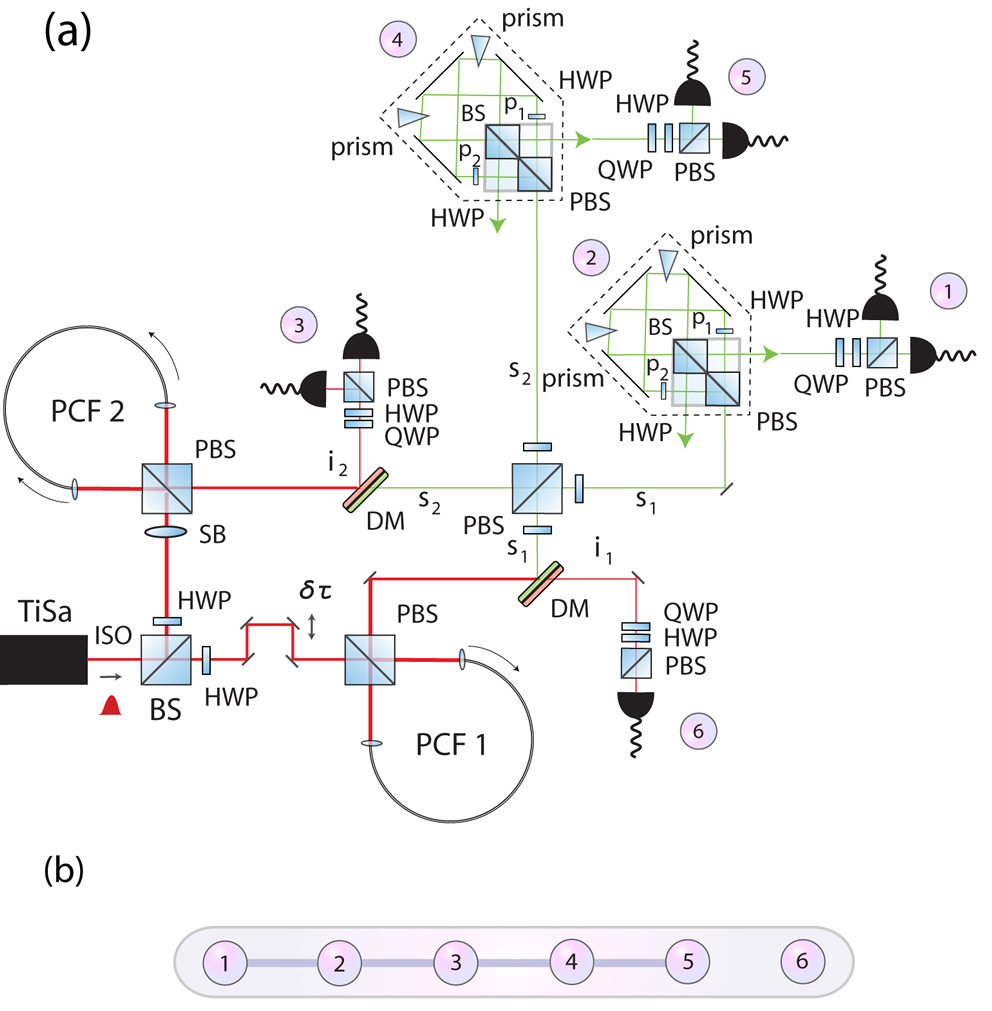}}
\caption{Experimental setup. {\bf (a):} Two photonic crystal fibres produce photon pairs which are fused using a polarizing beamsplitter (PBS) to generate the five-qubit entangled cluster state plus additional qubit 6 shown in (b). The cluster state consists of three polarization qubits, 1, 3 and 5 ($s_1$, $i_2$ and $s_2$). The paths of photons $s_1$ and $s_2$ represent qubits 2 and 4 respectively. The algorithm is executed by measuring the path qubits in the $Z$ or $Y$ bases depending on the oracle's black box using a Sagnac configuration (dashed regions). The output of the algorithm resides on qubits 1 and 5, and is obtained via polarization measurements. The setup is based on one recently used to generate a quantum error correction graph code~\cite{Bell14}, the main differences here being the use of an additional photon (qubit 6) and the waveplate configuration used to generate the different entangled resource. {\bf (b):} Cluster state generated by the setup. Edges correspond to controlled-phase operations, $CZ={\rm diag}\{ 1,1,1,-1 \}$, applied to qubits (the vertices) initialized in the state $|+ \rangle$.}
\label{setup}
\end{figure} 

The quantum algorithm used to solve SP considers the oracle as a black box (BB) implementing the following operation $U: \ket{x}\ket{z} \mapsto \ket{x}\ket{z \oplus f(x)}$, with $\ket{x}$ the query register and $\ket{z}$ an ancilla. The oracle is queried with a superposition of all inputs $\ket{x}$, and the ancilla state is $\ket{0}^{\otimes m}$ (where $\{\ket{0},\ket{1} \}$ is the qubit computational basis):$2^{-n/2}\sum_{x=0}^{2^n-1}\ket{x}\ket{0}^{\otimes m}$, which it transforms into $2^{-n/2}\sum_{x=0}^{2^n-1}\ket{x}\ket{f(x)}=2^{-n/2}(\ket{x_0}+\ket{x_0 \oplus s})\ket{f(x_0)}+2^{-n/2}(\ket{x_1}+\ket{x_1 \oplus s})\ket{f(x_1)}+\dots$.

Hadamard rotations $H=(X + Z)/\sqrt{2}$ are then applied to the oracle's output query qubits ($X,~Y$ and $Z$ are the Pauli matrices). Taking the first term as an example we have $(\ket{x_0}+\ket{x_0 \oplus s}) \to \sum_{y=0}^{2^n-1}[(-1)^{x_o \cdot y}+(-1)^{(x_o\oplus s) \cdot y}]\ket{y}$, where $\cdot$ is the bitwise inner product. Using the relation $(x_0 \oplus s)\cdot y=(x_0 \cdot y)\oplus (s\cdot y)$ we have that all terms with $(s\cdot y)=1$ interfere and cancel, leaving terms with $(s \cdot y)=0$ only. This cancellation occurs for the remaining terms involving $x_1, x_2, \dots x_{2^{n}-1}$. Thus measuring the query qubits gives a state $\ket{y}$ where $s \cdot y = 0$. By running the algorithm until $n-1$ linearly independent binary vectors $y_i$ are obtained (which occurs with $p_s \ge 1/4$ for $\geq n$ repetitions~\cite{Simon}) one can solve the list of $s \cdot y_i$'s to obtain $s$, in the case $f$ is $2-1$. If $f$ is $1-1$, a uniform distribution is found for the $y_i$ outcomes.

{\it Experimental implementation.-} The setup we use to demonstrate the algorithm is shown in Fig.~\ref{setup}~(a). Recently part of this setup was used to demonstrate a quantum error correction code using a 2D graph state~\cite{Bell14}. Here, the setup has been modified using a number of additional components in order to generate a different multipartite entangled state, a linear 1D cluster state. Using this different entangled state we are then able to demonstrate the quantum algorithm for SP. The ability to carry out a range of different protocols using cluster and graph states in this context shows their great flexibility for quantum information processing tasks~\cite{develop}.

In the setup a Ti:Sapphire laser emits 8nm pulses at a wavelength of $724.5$~nm with a repetition rate of 80~MHz, which are filtered to 1~nm. The pulses are split at a 50:50 beamsplitter and used to pump two birefringent photonic crystal fibre (PCF) sources. After filtering, attenuation, and coupling into the fibres, approximately 6mW/9mW is used to pump the first/second source. The PCF sources produce correlated pairs of photons via spontaneous four-wave mixing (FWM) at a signal and idler wavelength of $626$~nm and $860$~nm respectively~\cite{Fulconis3}. Here, the advantages of using FWM in a PCF to generate correlated photons compared to spontaneous parametric down-conversion in bulk crystals, such as BBO, include the ability to achieve pure state phase-matching~\cite{Halder}, as well as improved collection efficiencies and a lower pump power requirement~\cite{Fulconis3}.  Each source is in a Sagnac loop around a polarizing beamsplitter (PBS). In the first source the polarization of the pump is set to horizontal by a half-wave plate (HWP) and produces the state $\ket{H}_{i_1}\ket{H}_{s_1}$~\cite{Halder}. The photon pairs are separated into different paths using a dichroic mirror (DM) and the signal polarization is rotated to $\ket{+}$ using a HWP, where $\ket{\pm}=\frac{1}{\sqrt{2}}(\ket{H}\pm\ket{V})$. The detected rate of photon pairs in the state $\ket{H}_{i_1}\ket{+}_{s_1}$ is $\sim$$9,000$ per second, and the measured lumped efficiencies for the signal and idler paths are $\sim$$8\%$ and $\sim$$12\%$ respectively. The second PCF is pumped with diagonal polarization which produces the state $\frac{1}{\sqrt{2}}(\ket{H}\ket{H}+e^{i\theta}\ket{V}\ket{V})_{i_2s_2}$~\cite{Fulconis3}. A Soleil Babinet birefringent compensator (SB) placed before the loop is used to set the phase $\theta=0$, so that the Bell state $\ket{\phi^{+}}$ is produced. A DM separates the two wavelengths. The detected rate of photon pairs in this state is also $\sim$$9,000$ per second and the fidelity with respect to the ideal state is $\sim$$0.88$. When both PCFs simultaneously produce a photon pair, the combined state is $\ket{H}_{i_1}\ket{+}_{s_1} \frac{1}{\sqrt{2}}(\ket{H}_{i_2}\ket{H}_{s_2}+\ket{V}_{i_2}\ket{V}_{s_2})$. A tunable filter window set to $4$~nm bandwidth at $860$~nm is applied to the idlers. The idler modes are collected into single-mode fibres and used with coincidence detections at the signal modes to trigger the event in which four photons are generated.
 
The signal photons from each PCF are then fused using a PBS to make the three-photon linear cluster state $\frac{1}{\sqrt{2}}(\ket{+H+}+\ket{-V-})_{s_1\,i_2\,s_2}$~\cite{Bell14}. Here, the signal photons pass through $40$~nm bandwidth filters. These filters are used only to remove any remaining light coming from the bright pump beam, as the intrinsically pure state phase-matching from the FWM process ensures that narrow filtering is unnecessary for the signal photons, which have a bandwidth of $0.3$~nm~\cite{Halder}. Further details about the spectrum of the signal and idler photons can be found in Refs.~\cite{Fulconis3,Halder}. The coherence length of the signal photons interfering is $1$~mm~\cite{Bell14}. After the PBS part of the fusion, HWPs set at $45^{\circ}$ apply Hadamard rotations to the polarization states of the signal photons. The remaining idler photon $i_1$ is used as an additional qubit in the algorithm. Both signal photons are expanded into two qubits via a Sagnac configuration~\cite{Pan} (dashed boxes in Fig.~\ref{setup}~(a)). For the signal mode $s_1$, a PBS applies the transformations $\ket{H}_{s_{1}} \to \ket{H}_{s_{1}}\ket{p_1}_{s_{1}}$ and $\ket{V}_{s_{1}} \to \ket{V}_{s_{1}}\ket{p_2}_{s_{1}}$, where $\ket{p_{1(2)}}_{s_{1}}$ corresponds to the photon in path 1 (2). HWPs set at $45^{\circ}$ then apply Hadamard rotations to the polarization states in modes $p_1$ and $p_2$ to give $\ket{+}_{s_{1}}\ket{p_1}_{s_{1}}$ and $\ket{-}_{s_{1}}\ket{p_2}_{s_{1}}$ respectively. Applying the same transformations to $s_2$ we have
\bqa
&& \hskip-0.4cm\ket{\psi_\ell}=\frac{1}{2\sqrt{2}}\big[ \big(\ket{+}\ket{0}+\ket{-}\ket{1}\big)\ket{0}\big(\ket{0}\ket{+}+\ket{1}\ket{-}\big)+ \nonumber \\
&& \hskip0.3cm \big(\ket{+}\ket{0}-\ket{-}\ket{1}\big)\ket{1}\big(\ket{0}\ket{+}-\ket{1}\ket{-}\big)\big]_{12345}\ket{+}_6 
\eqa
where the polarization of photon $s_1$ represents qubit 1 ($\ket{0/1} \leftrightarrow \ket{H/V}$) and its path is qubit 2 ($\ket{0/1} \leftrightarrow \ket{p_1/p_2}$), and similarly for photon $s_2$, whose polarization represents qubit 5 and its path is qubit 4. The polarization of photon $i_{1(2)}$ is qubit 6 (3). The Hadamard basis $\ket{+/-}\leftrightarrow \ket{H/V}$ is used for qubit 6. The state $\ket{\psi_\ell}$ is a five-qubit linear cluster state with an additional qubit, $\ket{\psi_\ell}=\ket{\phi_{\cal C}}_{12345}\ket{+}_6$, as shown in Fig.~\ref{setup}~(b). The state is generated with a rate of $\sim$$0.25$ per second.
\begin{figure*}[t]
\centerline{\includegraphics[width=17.8cm]{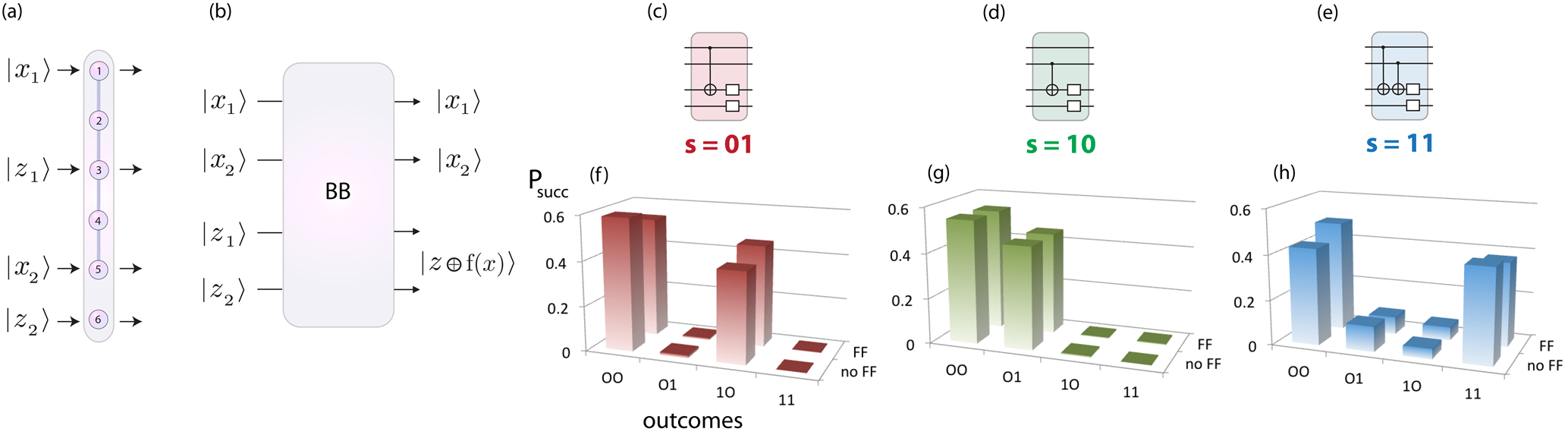}}
\caption{Black box (BB) circuit diagrams for SP$_{22}$ and experimental results. {\bf (a):} The cluster state resource used with additional qubit 6. {\bf (b):} General scenario for the oracle's BB, showing the inputs and outputs (reordered). {\bf (c)-(e):} Circuits corresponding to $2-1$ functions with $s=01$, $10$ and $11$ (the symbol $\Box$ corresponds to either a $\openone$ or $X$ operation). See Tab.~\ref{tab1} for values of $f(x)$ in each of these cases. {\bf (f)-(h):} Success probabilities measured in our experiment. Ideally the probabilities are equally split between outcomes $y_i=00$ and $10$ for $s=01$ (panel (f)), 00 and 01 for $s=10$ (panel (g)), and 00 and 11 for $s=11$ (panel (h)).}
\label{blackbox}
\end{figure*}

To measure the polarization qubits for the algorithm, each mode contains an analysis section made up of a quarter waveplate (QWP), HWP and PBS, allowing measurements in the $X$, $Y$, and $Z$ bases~\cite{James}. For the path qubits, $Z$ measurements are performed by blocking path $p_1$ or $p_2$ before the beamsplitter and measuring the relative populations~\cite{Pan}. For $X$ measurements, the paths are combined at the beamsplitter, which applies the transformation $\ket{p_1} \to \ket{+}$ and $\ket{p_2} \to \ket{-}$, with the output ports giving the relative populations. Instead of monitoring both output ports we modify the phase of one path relative to the other in order to swap the relative populations~\cite{Pan}. This allows measurements in the $Y$ basis also. For detecting the photons we use avalanche photodiodes and a coincidence counter monitors the 8 possible four-fold detections corresponding to one photon in each mode~\cite{Nock}. 

Before carrying out one-way QC on the cluster state, we characterize it, checking for the presence of genuine multipartite entanglement (GME) to ensure all photons are involved in the state generation. To do this, we measure the expectation value of the two setting witness, ${\cal W}_2$~\cite{TothGuhne},
which if negative detects the presence of GME. The witness is evaluated using the local measurements $XZXZX$ and $ZXZXZ$, and we find $\langle {\cal W}_2 \rangle=-0.12 \pm 0.02$, clearly showing the presence of GME. The error has been calculated using maximum likelihood estimation and a Monte Carlo method with Poissonian noise on the count statistics, which is the dominant source of error in our photonic experiment~\cite{James}. We also obtain the fidelity of the experimental cluster state with respect to the ideal state using seventeen measurement bases~\cite{supp} and find a fidelity of $F=0.70 \pm 0.01$.


With the cluster state characterized we implement the quantum algorithm. The action of the oracle is known as a promise problem~\cite{nielsenchuang}. In order to implement all the configurations that it might perform in an $n=m=2$ version of SP (SP$_{22}$), we must be able to construct them using a combination of quantum gates. There are a total of fifteen different oracle black boxes (BBs) for SP$_{22}$~\cite{supp}. However, in order to demonstrate the speedup achieved by the quantum algorithm it is not necessary to implement all BBs: the gap between the number of classical and quantum queries required to solve the problem is small for low $n$ and for $n=2$ there is no speedup if $1-1$ functions are included. SP stills applies to the case with only $2-1$ functions and a speedup exists for all $n\ge 2$~\cite{Simon}. In Fig.~\ref{blackbox}~(c), (d) and (e), we identify three BBs for $f$ as $2-1$ in terms of their equivalent quantum network, covering all periods $s=01$, 10 and 11 respectively. In order to carry out the algorithm using the necessary logic quantum gates, the five-qubit cluster state shown in Fig.~\ref{blackbox}~{\bf (a)} is used, where one-way QC is carried out by performing a program of measurements. No adjustment to the resource is necessary, each BB corresponds to a different program.

For cluster states, two types of measurements allow one-way QC to be performed: (i) Measuring a qubit $j$ in the computational basis allows it to be disentangled and removed from the cluster, leaving a smaller cluster of the remaining qubits, and (ii) In order to perform QC, qubits must be measured in the equatorial basis $B_j(\alpha)=\{ \ket{\alpha_+}_j,\ket{\alpha_-}_j \}$, where $\ket{\alpha_{\pm}}_j=(\ket{0}\pm e^{i \alpha}\ket{1})_j/\sqrt{2}$ ($\alpha\!\in\!{\mathbb R}$). Choosing the basis determines the rotation $R_z(\alpha)={\rm exp}(-i \alpha \sigma_z/2)$, which is followed by a Hadamard operation being simulated on a logical qubit in the cluster residing on qubit $j$~\cite{clusterback}. Using the cluster state generated, the input states corresponding to $\ket{x}=\ket{x_1}\ket{x_2}=\ket{+}\ket{+}$ are naturally encoded on qubits 1 and 5. The states $\ket{z_1}\ket{z_2}=\ket{0}\ket{0}$ are encoded on qubits 3 and 6, with the Hadamard operations from the BBs automatically applied before a particular measurement program begins. Thus, the state $\ket{x_1}\ket{z_1}\ket{x_2}\ket{z_2}\equiv\ket{+}(H\ket{0})\ket{+}(H\ket{0})$ resides on the logical input register of the resource $\ket{\psi_\ell}$. Qubits 2 and 4 play a pivotal role for the oracle by allowing it to apply (or not apply) two-qubit gates between the logical input states $\ket{x_1}$ and $\ket{z_1}$, and $\ket{x_2}$ and $\ket{z_1}$. For each BB, measuring a qubit in the computational basis prevents any two-qubit gate from being applied between its neighboring qubits. For example, in the BB of Fig.~\ref{blackbox}~(c), the oracle can measure qubit 4 in the computational basis, removing it from the cluster and leaving it with the ability to perform only a two-qubit gate between $\ket{x_1}$ and $\ket{z_1}$. When the oracle measures qubit 2 in $B(\pi/2)$ this enables it to apply the gate $(R_z(\pi/2) \otimes R_z(\pi/2)){\sf CZ}$ between $\ket{x_1}$ and $\ket{z_1}$~\cite{Tame3}, where ${\sf CZ}={\rm diag}(1,1,1,-1)$. This gives the computation $\ket{x_1}\ket{z_1}\ket{x_2}\ket{z_2} \to  [R_z(\pi/2) \otimes R_z(\pi/2) \otimes \openone \otimes \openone][{\sf CZ} \otimes \openone \otimes \openone][\openone \otimes H \otimes \openone \otimes H]\ket{+}\ket{0}\ket{+}\ket{0} \equiv {\sf CNOT}\otimes \openone \otimes \openone \ket{+}\ket{0}\ket{+}\ket{0}$, up to local rotations $[R_z(-\pi/2)\otimes H\,R_z(-\pi/2)\otimes \openone \otimes H]_{1356}$ incorporated into a `feed-forward' (FF) stage. These FF rotations are realised in the experiment by modifying the basis of the measurements of the corresponding qubits - a standard procedure in one-way QC where a local unitary operation before a measurement is equivalent to a basis change of the measurement itself~\cite{onewaye}. The above combination of logic gates and FF corresponds to the required circuit for the BB of Fig.~\ref{blackbox}~(c). For measurement outcomes $s_2=s_4=0$ the final state (with FF applied) is
\be
\ket{\psi_\ell'}=\frac{1}{\sqrt{2}}\big(\ket{0}_1\ket{-}_3+\ket{1}_1\ket{+}_3 \big)\ket{0}_5\ket{0}_6, 
\ee
which gives the outcomes for the query qubits (when measured in the computational basis) of $y_i=s_1s_5$ equal to $00$ or $10$, as expected. For the other measurement outcomes of qubits 2 and 4 one applies FF operations to qubits 1 and 5 given in Table 1 by incorporating them into the measurement basis. Full details of the evolution of the cluster state resource during these steps can be found in Ref.~\cite{supp}.
\begin{table}[t]
\begin{ruledtabular} 
\begin{tabular}{| >{\centering\arraybackslash} b{2cm}| >{\centering\arraybackslash} b{2cm}| >{\centering\arraybackslash} b{2cm}| >{\centering\arraybackslash} b{2cm}|}
\hline
$f(x)$ & $s=01$ & $s=10$ & $s=11$ \\ 
\hline
$f(00)$ & 00 & 00 & 00 \\
$f(01)$ & 00 & 10 & 10 \\
$f(10)$ & 10 & 00 & 10 \\
$f(11)$ & 10 & 10 & 00 \\
\hline \hline
${\cal M}_2$ & $B(\pi/2)$ & $\ket{0/1}$ & $B(\pi/2)$ \\
${\cal M}_4$ & $\ket{0/1}$ & $B(\pi/2)$ & $B(\pi/2)$ \\
\hline \hline
FF$_1$ & $\chi^{20}$ & $\zeta^{20}$ & $\chi^{20}$ \\
FF$_3$ & $\chi^{24}$ & $\chi^{24}$ & $\tilde{\chi}^{24}$ \\
FF$_5$ & $\zeta^{04}$ & $\chi^{04}$ & $\chi^{04}$ \\
FF$_6$ & H & H & H \\
\hline
\end{tabular}
\end{ruledtabular}
\caption{ BB function outputs $f(x)$ for SP$_{22}$ (rows 1-4, the $y_i$ outputs from the quantum algorithm are given later in the main text), measurement program ${\cal M}_i$ for qubit $i$ of the cluster state $\ket{\phi_{\cal C}}$ (rows 5,6) and FF operations (rows 7-10) for each period $s$. The notation $\ket{0/1}$ corresponds to a measurement in the computational basis with $\chi^{ij}=X^{s_i+s_j}HR_z(-\pi/2)$, $\tilde{\chi}^{ij}=X^{s_i+s_j}HR_z(-\pi)$ and $\zeta^{ij}=HX^{s_i+s_j}$. Here, $s_k$ is the measurement outcome of qubit $k$ (with $s_0=0$). For each period, $s$, there are an additional three function outputs, obtained by applying the combination $\openone \otimes X$, $X \otimes \openone$ and $X \otimes X$ to ancilla qubits $z_1$ and $z_2$ (see $\Box$'s in Fig.~\ref{blackbox}).}
\label{tab1}
\end{table}

The same basic rules can be applied for all the BBs. Table~\ref{tab1} provides the measurement programs for each BB. Here, the final Hadamards for the query qubits after the BB's (before they are measured) are also added to the FF stage, allowing the algorithm for SP$_{22}$ to be implemented. The measurements and outcomes of qubits $1$, $3$, $5$ and $6$ constitute the algorithm (only query qubits 1 and 5 need to be measured to obtain $y_i$). Additions to the FF stages, and measurements of qubits 2 and 4 are viewed as being carried out by the oracle~\cite{Tame1}. 

The results of the experiment are shown in Fig.~\ref{blackbox}~(f)-(h), where we display the average success probability of obtaining the different logic outcomes of the query qubits for each BB function shown in (c)-(e). For the BB with $s=01$, the success probabilty should be split equally between $y_i=00$ and $y_i=10$, as $s \cdot y_i = 0$ and $\cdot$ is the bitwise inner product. We find $p_{00}=0.54 \pm 0.02$ and $p_{10}=0.45 \pm 0.02$ as shown in Fig.~\ref{blackbox}~(f). For the BB with $s=10$ ($s=11$), the success probabilty should be split equally between $y_i=00$ and $y_i=01$ ($y_i=11$) as $s \cdot 00=s \cdot 01 = 0$ ($s \cdot 00=s \cdot 11 = 0$). We find $p_{00}=0.54 \pm 0.02$ and $p_{01}=0.45 \pm 0.01$ ($p_{00}=0.49 \pm 0.02$ and $p_{11}=0.37 \pm 0.01$) as shown in Fig.~\ref{blackbox}~(g) ((h)). In Fig.~\ref{blackbox}~(f)-(h) we include the no-FF ($s_i=0~\forall i$) and FF cases for the algorithm~\cite{footFF}. Note that we have repeated the algorithm a number of times to obtain the success probabilities. However, on average only~$\sim2$ runs are sufficient in order to obtain an outcome other than 00. This is in contrast to the classical scenario which requires on average 8/3 runs to determine the period~\cite{supp}. While this gap between the quantum and classical runtime is small in the two-qubit version, it scales exponentially with the size of the input register. Our results provide the first experimental evidence of the existence of this gap. We have briefly analyzed the resources required for demonstrating $n$-qubit versions of the algorithm for SP and found that the minimal graph state for performing SP$_{nn}$ contains $n^2+n+1$ qubits and $2n^2-2n+2$ edges. This resource scales polynomially with $n$~\cite{supp}. The six-qubit resource used here for SP$_{22}$ is a special case excluding $1-1$ functions.

{\it Remarks.-} We have reported the first experimental realization of a two-qubit version of the algorithm for Simon's Problem, a black box problem, showing the first hint of an exponential gap existing between the classical and quantum runtime. The agreement between the experimental data and theory is excellent and limited only by the overall quality of the resource. The experiment has been performed in a photonic system, which due to the strong potential of using photonics for advanced quantum information processing, makes our scheme ideal for future probing of the boundary between classical and quantum efficiency in computing algorithms. Subsequent work on applying the techniques described here to other quantum algorithms may further stimulate one-way QC with minimal resources and their expansion to full-scale quantum information processing.

{\it Acknowledgments.-} We thank M. Paternostro for helpful discussions, and support from EU project 600838 QWAD, UK EPSRC EP/K034480/1 and ERC grant 247462 QUOWSS.



\newpage

\setcounter{equation}{0}
\setcounter{figure}{0}

\renewcommand{\figurename}{FIG. S}
\renewcommand{\thefigure}{\arabic{figure}}

\begin{widetext}


\section{Supplemental Material}

\subsection{Fidelity of cluster state resource}
To obtain the fidelity of the five-qubit cluster state we decompose the fidelity operator into a summation of products of Pauli matrices as
\bqa
F&=&\ketbra{\phi_C}{\phi_C}=\frac{1}{32}(1+IIIZX+IIZXZ+IIZYY+IZXIX+IZXZI-IZYXY+IZYYZ+XIXIX+XIXZI-XIYXY\\
&&\hskip1.8cm+XIYYZ+XZIII+XZIZX+XZZXZ+XZZYY-YXXXY+YXXYZ-YXYIX-YXYZI+YYIXZ \nonumber \\
&&\hskip1cm +YYIYY+YYZII +YYZZX+ZXIXZ+ZXIYY+ZXZII+ZXZZX+ZYXXY-ZYXYZ+ZYYIX+ZYYZI). \nonumber
\eqa
Calculating the expectation value of this operator requires 17 unique measurement bases: $XZZXZ$, $XZZYY$, $YXXXY$, $YXXYZ$, $YYZZX$, $ZXZZX$, $ZYXXY$, $ZYXYZ$, $XZXZX$, $XZYXY$, $XZYYZ$, $YXYZX$, $YYIXZ$, $YYIYY$, $ZXIXZ$, $ZXIYY$ and $ZYYZX$.

\subsection{Evolution of underlying cluster state resource for the black box functions}

For the black box (BB) functions we have the starting resource state 
\be
\ket{\psi_\ell}=\frac{1}{2\sqrt{2}}\big[ \big(\ket{+}\ket{0}+\ket{-}\ket{1}\big)\ket{0}\big(\ket{0}\ket{+}+\ket{1}\ket{-}\big)+ \big(\ket{+}\ket{0}-\ket{-}\ket{1}\big)\ket{1}\big(\ket{0}\ket{+}-\ket{1}\ket{-}\big)\big]_{12345}\ket{+}_6.
\ee
The logical state $\ket{x_1}\ket{z_1}\ket{x_2}\ket{z_2}\equiv\ket{+}(H\ket{0})\ket{+}( H\ket{0})$ resides on the input register of the resource $\ket{\psi_\ell}$, where $\ket{x_1}$ resides on qubit 1, $\ket{x_2}$ on qubit 5, $\ket{z_1}$ on qubit 3 and $\ket{z_2}$ on qubit 6. 
\vskip0.5cm
For BB~(c), corresponding to $s=01$, the oracle measures qubit 4 in the computational basis, removing it from the cluster and leaving it with the ability to perform only a two-qubit gate between $\ket{x_1}$ and $\ket{z_1}$. In this case the resource state $\ket{\psi_\ell}$ becomes
\be
\ket{\psi_\ell'}=\frac{1}{2\sqrt{2}}\bigg[\bigg(\big(\ket{+}\ket{0}+\ket{-}\ket{1}\big)\ket{0}+(-1)^{s_4}\big(\ket{+}\ket{0}-\ket{-}\ket{1}\big)\ket{1}\bigg)_{123}\big(\ket{0}+(-1)^{s_4}\ket{1}\big)_5\ket{+}_6\bigg].
\ee
The oracle then measures qubit 2 in $B(\pi/2)$, which enables it to apply the gate $(R_z(\pi/2) \otimes R_z(\pi/2)){\sf CZ}$ between $\ket{x_1}$ and $\ket{z_1}$~\cite{STame1,STame2}, where ${\sf CZ}={\rm diag}(1,1,1,-1)$. This gives the computation $\ket{x_1}\ket{z_1}\ket{x_2}\ket{z_2} \to  [R_z(\pi/2) \otimes R_z(\pi/2) \otimes \openone \otimes \openone][{\sf CZ} \otimes \openone \otimes \openone][\openone \otimes H \otimes \openone \otimes H]\ket{+}\ket{0}\ket{+}\ket{0}$. The resource state becomes
\be
\ket{\psi_\ell''}=\frac{1}{2\sqrt{2}}\bigg[\bigg(\big(\ket{0}+i(-1)^{s_2}\ket{1}\big)\ket{0}+i(-1)^{s_2 \oplus s_4}\big(\ket{0}-i(-1)^{s_2}\ket{1}\big)\ket{1}\bigg)_{13}\big(\ket{0}+(-1)^{s_4}\ket{1}\big)_5\ket{+}_6\bigg].
\ee
By incorporating the local rotations $[R_z(-\pi/2)\otimes H\,R_z(-\pi/2)\otimes \openone \otimes H]_{1356}$ into the feed-forward (FF) stage~\cite{STame1}, this gives the logical operation ${\sf CNOT}\otimes \openone \otimes \openone \ket{+}\ket{0}\ket{+}\ket{0}$. By also including the final Hadamards on the query qubits (1 and 5) in the FF stage we have the output of the algorithm as the final state
\bqa
\ket{\psi_\ell'''}&=&\frac{1}{\sqrt{2}}\bigg[\bigg(\bigg(\frac{1}{2}\big(1-(-1)^{s_2}\big)\ket{0}+\frac{1}{2}\big(1+(-1)^{s_2}\big)\ket{1}\bigg)_1\ket{+}_3+(-1)^{s_2 \oplus s_4}\bigg(\frac{1}{2}\big(1+(-1)^{s_2}\big)\ket{0}+\frac{1}{2}\big(1-(-1)^{s_2}\big)\ket{1}\bigg)_1\ket{-}_3\bigg)\otimes \nonumber \\
&&\hskip9cm\bigg(\frac{1}{2}\big(1+(-1)^{s_4}\big)\ket{0}+\frac{1}{2}\big(1-(-1)^{s_4}\big)\ket{1}\bigg)_5\ket{0}_6 \bigg].
\eqa
For measurement outcomes $s_2=s_4=0$ the final state is
\be
\ket{\psi_\ell'''}=\frac{1}{\sqrt{2}}\bigg(\ket{0}_1\ket{-}_3+\ket{1}_1\ket{+}_3 \bigg)\ket{0}_5\ket{0}_6,
\ee
which gives the outcomes for the query qubits (when measured in the computational basis) of $y_i=s_1s_5$ equal to $00$ or $10$, as expected. For the other measurement outcomes of qubits 2 and 4 one applies the FF operations to qubits 1 and 5 given in Table 1 of the main text, which is achieved by modifying the measurement basis of the qubits.

For BB~(d) and BB~(e) similar steps can be derived as outlined above. For BB~(d) we have for measurement outcomes $s_2=s_4=0$ the final state
\be
\ket{\psi_\ell'''}=\ket{0}_1\frac{1}{\sqrt{2}}\bigg(\ket{+}_3\ket{1}_5+\ket{-}_3\ket{0}_5 \bigg)\ket{0}_6,
\ee
which gives the outcomes for the query qubits (when measured in the computational basis) of $y_i=s_1s_5$ equal to $00$ or $01$. For the other measurement outcomes of qubits 2 and 4 one applies the FF operations to qubits 1 and 5 given in Table 1 of the main text.

For BB~(e) we have for measurement outcomes $s_2=s_4=0$ the final state
\be
\ket{\psi_\ell'''}=\frac{1}{\sqrt{2}}\big( \ket{1}_1 \ket{+}_3 \ket{1}_5+ \ket{0}_1 \ket{-}_3 \ket{0}_5 \big)\ket{0}_6,
\ee
which gives the outcomes for the query qubits (when measured in the computational basis) of $y_i=s_1s_5$ equal to $00$ or $11$. For the other measurement outcomes of qubits 2 and 4 one applies the FF operations to qubits 1 and 5 given in Table 1 of the main text.

\subsection{All SP$_{22}$ functions using an 8-qubit cluster state resource}

Here we show how one can implement all the black box (BB) functions for the oracle in Simon's Problem (SP) using an 8-qubit cluster state. In Fig.~S~\ref{blackbox2} we show all possible oracles for an $n=m=2$ version of SP (SP$_{22}$) in terms of their equivalent quantum network. One can see that all fifteen oracle black boxes (BB(a)-(o)) implement their respective oracle operation given in Tab.~\ref{tab2}. In order to carry out the algorithm for SP$_{22}$ which uses these quantum gates, the eight-qubit cluster state shown in Fig.~S~\ref{resource}~(a) can be used, where one-way QC is carried out by performing a correct program of measurements. No adjustment to the resource is necessary; each BB corresponds to a different measurement program on the same resource.
\begin{figure*}[b]
\includegraphics[width=16cm]{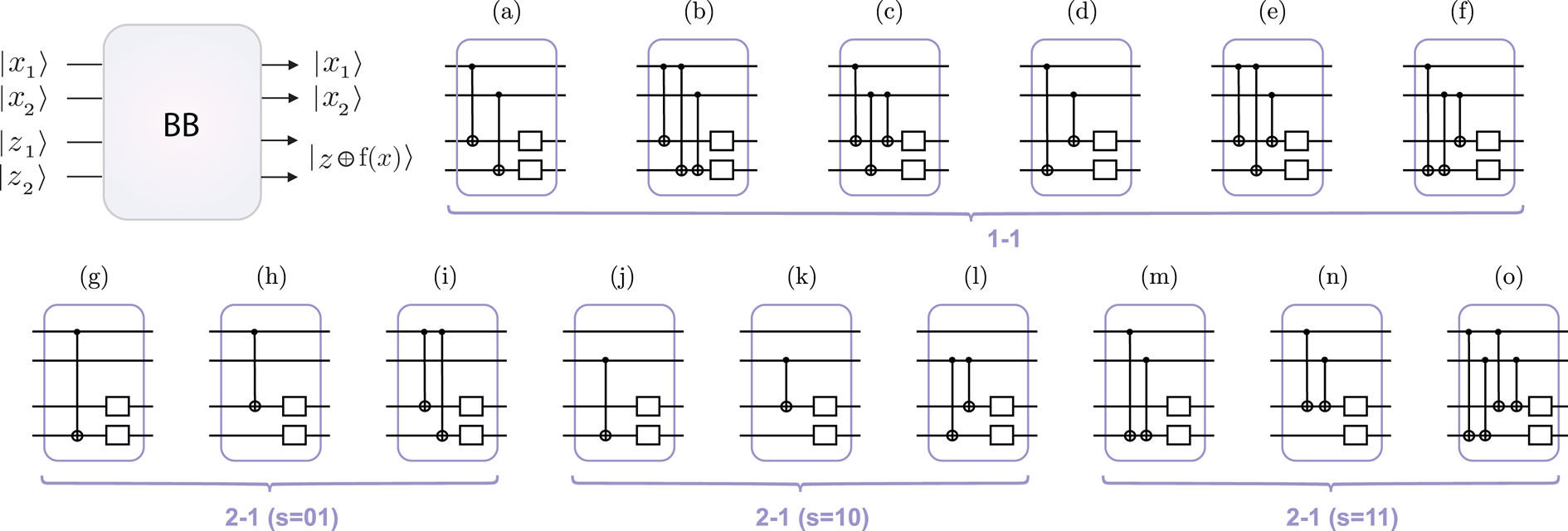}
\caption{All black box circuit diagrams for SP$_{22}$. Circuits (a)-(f) correspond to $1-1$ functions while (g)-(o) correspond to $2-1$ functions. Here $\Box$ corresponds to either a $\openone$ or $\sigma_x$ operation. See Tab.~\ref{tab2} for individual values of $f(x)$ in each case.}
\label{blackbox2}
\end{figure*}

We now describe how the algorithm is implemented on the eight-qubit cluster state and then show how it is compacted to the resource of six qubits realized in our experiment. The 8-qubit cluster resource shown in Fig.~S~\ref{resource}~(a), which we denote as $\ket{\Phi_{\cal C}}$, is given by
\bqa
\ket{\Phi_{\cal C}}&=&\frac{1}{4}[(\ket{0+0+0}+\ket{0-1-0})(\ket{0++}+\ket{1--})+(\ket{0+0-1}+\ket{0-1+1})(\ket{1+-}+\ket{0-+}) \nonumber \\
& &  (\ket{1+1+1}+\ket{1-0-1})(\ket{1++}+\ket{0--})+(\ket{1+1-0}+\ket{1-0+0})(\ket{0+-}+\ket{1-+})]_{1 \to 8}.
\eqa
The input states corresponding to $\ket{x}=\ket{x_1}\ket{x_2}=\ket{+}\ket{+}$ are encoded on qubits 1 and 5. The states $\ket{z_1}\ket{z_2}=\ket{0}\ket{0}$ are encoded on qubits 3 and 6, with Hadamard operations from the BB's automatically applied before the measurement program begins. Thus, the state $\ket{x_1}\ket{z_1}\ket{x_2}\ket{z_2}\equiv\ket{+}(H\ket{0})\ket{+}( H\ket{0})$ resides on the logical input register of the cluster $\ket{\Phi_C}$. 
Qubits 2, 4, 7 and 8 play the pivotal role of the oracle by performing two-qubit gates between the logical input states $\ket{x_1}$ and $\ket{z_1}$ (or $\ket{z_2}$) and $\ket{x_2}$ and $\ket{z_1}$ (or $\ket{z_2}$). For each BB, measuring a qubit in the computational basis disentangles it from the cluster and prevents any two-qubit gate from being applied between its neighboring qubits. For example, in BB(a), the oracle can measure qubits 4 and 8, leaving it with the ability to perform only a two-qubit gate between $\ket{x_1}$ and $\ket{z_1}$ together with one between $\ket{x_2}$ and $\ket{z_2}$. When the oracle measures qubits 2 and 7 both in the $B(\pi/2)$ basis this enables it to apply the gate $(R_z(\pi/2) \otimes R_z(\pi/2)){\sf CZ}$ between $\ket{x_1}$ and $\ket{z_1}$, together with the same gate between $\ket{x_2}$ and $\ket{z_2}$ (see Tame {\it et al.} in~\cite{STame1,STame2}), where ${\sf CZ}$ shifts the relative phase of the state $\ket{1}\ket{1}$ by $\pi$ with respect to the rest of the computational basis states. This gives the computation $\ket{x_1}\ket{z_1}\ket{x_2}\ket{z_2} \to  [R_z(\pi/2) \otimes R_z(\pi/2) \otimes R_z(\pi/2) \otimes R_z(\pi/2)][{\sf CZ} \otimes {\sf CZ}][\openone \otimes H \otimes \openone \otimes H]\ket{+}\ket{0}\ket{+}\ket{0} \equiv {\sf CNOT}\otimes{\sf CNOT} \ket{+}\ket{0}\ket{+}\ket{0}$, up to local rotations $[R_z(-\pi/2)\otimes H\,R_z(-\pi/2)\otimes R_z(-\pi/2)\otimes H\,R_z(-\pi/2)]_{1356}$ incorporated into the feed-forward (FF) stage~\cite{STame1}. This corresponds to the required circuit for BB(a). The same basic rules described above can be applied to all the BB's. The circuits can always be decomposed into ${\sf CZ}$'s, with any excess Hadamards applied at the FF stages. Tab.~\ref{tab2} provides the measurement programs for each BB. Here, the final Hadamards required after the BB's (before the query qubits are measured) are also added to the FF stage, thus allowing the full algorithm for SP$_{22}$ to be implemented on an eight-qubit cluster state. The measurements and outcomes of qubits $1$, $3$, $5$ and $6$ constitute the algorithm, although only the query qubits 1 and 5 need actually be measured to obtain the value $y$. The additions to the FF stages described above, together with the measurements of qubits 2, 4, 7 and 8 should be viewed as being carried out entirely by the oracle. 
\begin{table*}[t]
\begin{ruledtabular} 
\begin{tabular}{|c|c|c|c|c|c|c|c|c|c|c|c|c|c|c|c|}
\hline
\phantom{p}$f(x)$ & $a$ & $b$ & $c$ & $d$ & $e$ & $f$ & $g$ & $h$ & $i$ & $j$ & $k$ & $l$ & $m$ & $n$ & $o$ \\ \hline
\phantom{p}$f(00)$ & 00 & 00 & 00 & 00 & 00 & 00 & 00 & 00 & 00 & 00 & 00 & 00 & 00 & 00 & 00 \\
\phantom{p}$f(01)$ & 01 & 01 & 11 & 10 & 10 & 11 & 00 & 00 & 00 & 01 & 10 & 11 & 01 & 10 & 11 \\
\phantom{p}$f(10)$ & 10 & 11 & 10 & 01 & 11 & 01 & 01 & 10 & 11 & 00 & 00 & 00 & 01 & 10 & 11 \\
\phantom{p}$f(11)$ & 11 & 10 & 01 & 11 & 01 & 10 & 01 & 10 & 11 & 01 & 10 & 11 & 00 & 00 & 00 \\
\hline \hline
${\cal M}_2$ & $B(\frac{\pi}{2})$ & $B(\frac{\pi}{2})$ & $B(\frac{\pi}{2})$ & $\ket{0/1}$ & $B(\frac{\pi}{2})$ & $\ket{0/1}$ & $\ket{0/1}$ & $B(\frac{\pi}{2})$ & $B(\frac{\pi}{2})$ & $\ket{0/1}$ & $\ket{0/1}$ & $\ket{0/1}$ & $\ket{0/1}$ & $B(\frac{\pi}{2})$ & $B(\frac{\pi}{2})$ \\
${\cal M}_4$ & $\ket{0/1}$ & $\ket{0/1}$ & $B(\frac{\pi}{2})$ & $B(\frac{\pi}{2})$ & $B(\frac{\pi}{2})$ & $B(\frac{\pi}{2})$ & $\ket{0/1}$ & $\ket{0/1}$ & $\ket{0/1}$ & $\ket{0/1}$ & $B(\frac{\pi}{2})$ & $B(\frac{\pi}{2})$ & $\ket{0/1}$ & $B(\frac{\pi}{2})$ & $B(\frac{\pi}{2})$ \\
${\cal M}_7$ & $B(\frac{\pi}{2})$ & $B(\frac{\pi}{2})$ & $B(\frac{\pi}{2})$ & $\ket{0/1}$ & $\ket{0/1}$ & $B(\frac{\pi}{2})$ & $\ket{0/1}$ & $\ket{0/1}$ & $\ket{0/1}$ & $B(\frac{\pi}{2})$ & $\ket{0/1}$ & $B(\frac{\pi}{2})$ & $B(\frac{\pi}{2})$ & $\ket{0/1}$ & $B(\frac{\pi}{2})$ \\
${\cal M}_8$ & $\ket{0/1}$ & $B(\frac{\pi}{2})$ & $\ket{0/1}$ & $B(\frac{\pi}{2})$ & $B(\frac{\pi}{2})$ & $B(\frac{\pi}{2})$ & $B(\frac{\pi}{2})$ & $\ket{0/1}$ & $B(\frac{\pi}{2})$ & $\ket{0/1}$ & $\ket{0/1}$ & $\ket{0/1}$ & $B(\frac{\pi}{2})$ & $\ket{0/1}$ & $B(\frac{\pi}{2})$ \\
\hline \hline
FF$_1$ & $\chi^{28}$ & $\tilde{\chi}^{28}$ & $\chi^{28}$ & $\chi^{28}$ & $\tilde{\chi}^{28}$ & $\chi^{28}$ & $\chi^{28}$ & $\chi^{28}$ & $\tilde{\chi}^{28}$ & $\zeta^{28}$ & $\zeta^{28}$ & $\zeta^{28}$ & $\chi^{28}$ & $\chi^{28}$ & $\tilde{\chi}^{28}$ \\
FF$_3$ & $\chi^{24}$ & $\chi^{24}$ & $\tilde{\chi}^{24}$ & $\chi^{24}$ & $\tilde{\chi}^{24}$ & $\chi^{24}$ & $\zeta^{24}$ & $\chi^{24}$ & $\chi^{24}$ & $\zeta^{24}$ & $\chi^{24}$ & $\chi^{24}$ & $\chi^{24}$ & $\tilde{\chi}^{24}$ & $\tilde{\chi}^{24}$ \\
FF$_5$ & $\chi^{74}$ & $\chi^{74}$ & $\tilde{\chi}^{74}$ & $\chi^{74}$ & $\chi^{74}$ & $\tilde{\chi}^{74}$ & $\zeta^{74}$ & $\zeta^{74}$ &  $\zeta^{74}$ & $\chi^{74}$ & $\chi^{74}$ & $\tilde{\chi}^{74}$ & $\chi^{74}$ & $\chi^{74}$ & $\tilde{\chi}^{74}$ \\
FF$_6$ & $\chi^{87}$ & $\tilde{\chi}^{87}$ & $\chi^{87}$ & $\chi^{87}$ & $\chi^{87}$ & $\tilde{\chi}^{87}$ & $\chi^{87}$ & $\zeta^{87}$ & $\chi^{87}$ & $\chi^{87}$ & $\zeta^{87}$ & $\chi^{87}$ & $\tilde{\chi}^{87}$ & $\zeta^{87}$ & $\tilde{\chi}^{87}$ \\
\hline
\end{tabular}
\end{ruledtabular}
\caption{Black box function outputs for SP$_{22}$ together with the measurement program ${\cal M}_i$ for qubit $i$ of the cluster state $\ket{\Phi_C}$ and FF operations. The notation $\ket{0/1}$ corresponds to a measurement in the computational basis with $\chi^{ij}=\sigma_x^{s_i+s_j}HR_z(-\pi/2)$, $\tilde{\chi}^{ij}=\sigma_x^{s_i+s_j}HR_z(-\pi)$ and $\zeta^{ij}=H\sigma_x^{s_i+s_j}$. For each case denoted by a {\it letter} there are an additional three function outputs, obtained by applying the combination $\openone \otimes \sigma_x$, $\sigma_x \otimes \openone$ and $\sigma_x \otimes \sigma_x$ to the ancilla qubits $z_1$ and $z_2$ (see $\Box$'s in Fig.~S~\ref{blackbox2}).}
\label{tab2}
\end{table*} 

As mentioned in the main text, in order to demonstrate the speedup achieved by the quantum algorithm in the two-qubit case, it is not necessary to implement all of the functions in SP$_{22}$. If we consider only representatives of the functions, {\it i.e.} selected functions corresponding to each $s$, it is possible to show speedup with just a six-qubit cluster state. An important aid in the reduction of the resource is the observation that the gap between the number of classical queries and quantum queries required to solve the problem is small for low $n$. It turns out that for $n=2$ there is no speedup if $1-1$ functions are included as possible actions by the oracle. Fortunately, SP stills applies to the case where only $2-1$ functions are involved and a speedup exists for all $n\ge 2$~\cite{SSimon}. In Fig.~S~\ref{blackbox2}, we identify BBs (h), (k) and (n) for $f$ as $2-1$, covering all periods $s=01$, 10 and 11 respectively. These are the BBs used in our experiment. In these BB's, no two-qubit gate is present between $\ket{z_2}$ and any other logical qubit. By removing qubits 7 and 8 from the cluster we obtain the six-qubit cluster state shown in Fig.~S~\ref{resource}~(b). The measurement programs for this smaller resource follow those given in Tab.~\ref{tab2} for the eight-qubit case, where the values $s_7$ and $s_8$ corresponding to the measurement outcomes of qubits 7 and 8 should be disregarded. 

\subsection{Classical runtime for SP$_{22}$}

One can calculate the number of runs required classically for SP$_{22}$ by considering all possible function outputs from the BBs. For the limited set of BB's used in the experiment (representatives of the $2-1$ functions) and taking the inputs 00, 01, 10 and 11, we have from Tab.~\ref{tab2} the outputs of the query qubits given by 00, 00, 10 and 10 for $s=01$ (BB(h)), 00, 10, 00 and 10 for $s=10$ (BB(k)), and 00, 10, 10 and 00 for $s=11$ (BB(n)). The oracle also has the ability to apply bit flips to these outputs. By considering all $3 \times 4=12$ output permutations, one finds that if the oracle picks at random the period $s$ and the corresponding bit flips, the average number of queries needed to determine $s$ is $8/3$. 

\subsection{Scaling to SP$_{nn}$}

Here we comment briefly on extending the method proposed in the main text for performing $n$-qubit versions of the algorithm on cluster states for representatives of the oracle functions in SP$_{nn}$. Classically for a $1-1$ function, $n$ {\sf CNOT} gates can be used to directly copy each query bit $x_i$ ($i \in n$) to its corresponding ancilla bit $z_i$ (see Fig.~S~\ref{blackbox2}~(a)). With this, the oracle can implement $2^n$ different $1-1$ functions using combinations of bit-flip operations on the output ancilla bits. For $2-1$ functions, consider the case where the $i^{th}$ bit in $s$ has the value 1, {\it e.g.} $s=00\dots 1 \dots 00$. This can be implemented by again copying the contents of the query register to the ancilla register, except for the $i^{th}$ bit (see Fig.~S~\ref{blackbox2}~(h) ((j)) for $s=01$ ($s=10$)). Two different $x$ values, $a$ and $b$, the same in all bits except for the $i^{th}$ bit (thus satisfying $a = b \oplus s$), will then produce the same output $f(x)$. The overhead is $n-1$ {\sf CNOT}'s. Now consider that there are two bits in $s$ with the value 1 at positions $i$ and $j$, {\it e.g.} $s=00 \dots 1 \dots 1 \dots 00$. The value $x_i \oplus x_j$ must be the same for two different $x$ values $a$ and $b$ to give the same output $f(x)$. Two {\sf CNOT}'s, with controls on the $i^{th}$ and $j^{th}$ query bits and both targets on either the $i^{th}$ {\it or} $j^{th}$ ancilla bit (with the remaining bit left untouched), together with all other bits having {\sf CNOT} gates between the query and ancilla registers, can {\it check} that $x_i \oplus x_j$ is the same for $a$ and $b$ and implement the $2-1$ function (see Fig.~S~\ref{blackbox2}~ (m) or (n) for $s=11$). 
\begin{figure}[t]
\centerline{\includegraphics[width=9.0cm]{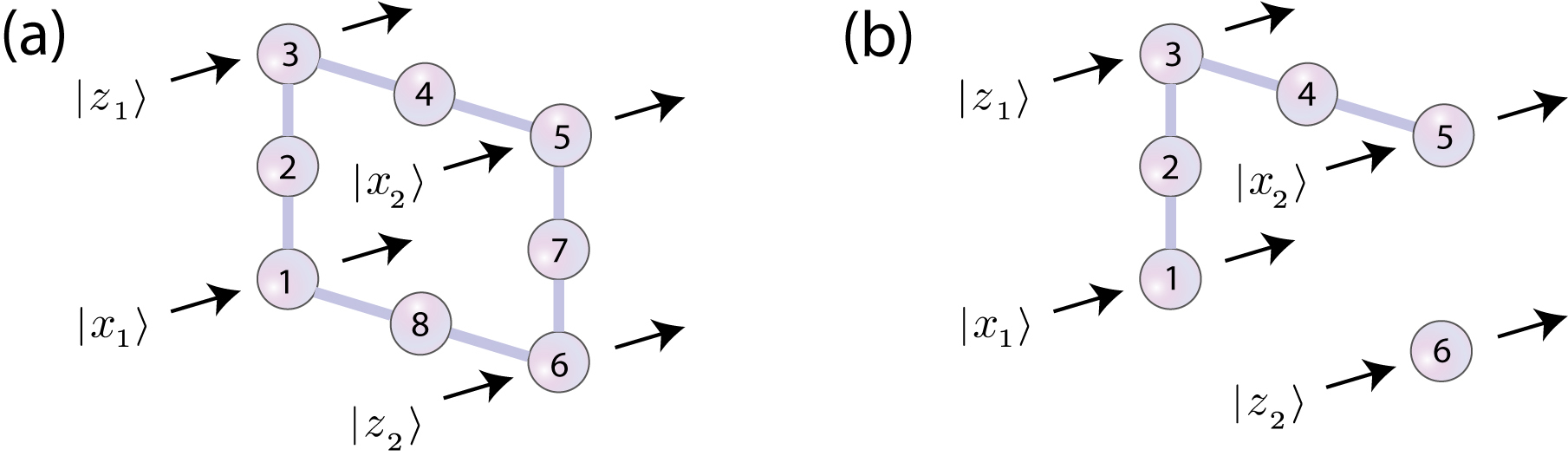}}
\caption{Cluster states for SP$_{22}$. {\bf (a)}: The eight-qubit cluster state $\ket{\Phi_{\cal C}}$ depicting the input/output logical qubits. {\bf (b)}: Six-qubit configuration for representatives of the functions as demonstrated in the experiment.}
\label{resource}
\end{figure}
\begin{figure}[t]
\centerline{\includegraphics[width=5.7cm]{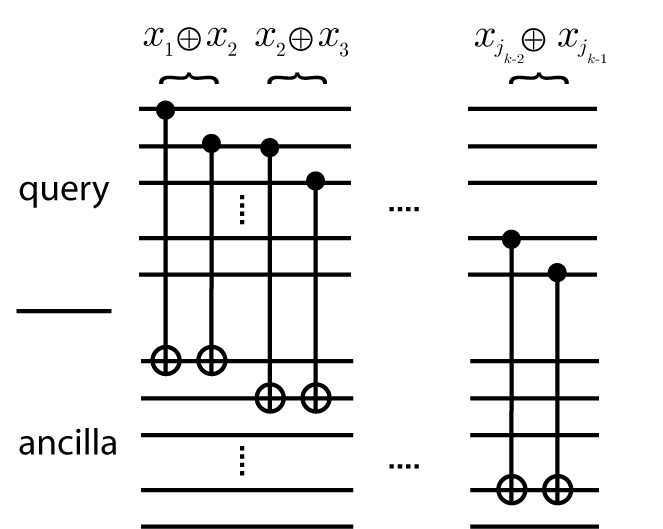}}
\caption{Building up to general $n$-qubit versions of the algorithm for SP. Pairwise {\it checks} for the $2-1$ function with $s=11\dots 11$ in the $n$-bit case, {\it i.e.} $k=n$.}
\label{resource2}
\end{figure}
In general, for $k$ bits with the value 1 in $s$, $x_i \oplus x_j$ {\it checks} (each using two {\sf CNOT}'s) are required on all query bits $\ell$, where $s_\ell=1$ in $s$, keeping the $i^{th}$ bit fixed as the first occurring 1 in $s$. However, each ancilla bit cannot be used more than once as a target for the check operations. The checks should therefore be made pairwise, {\it e.g.} $x_i \oplus x_{j_1}$, $x_{j_1} \oplus x_{j_2}$,\dots $x_{j_{k-2}} \oplus x_{j_{k-1}}$, as shown in Fig.~S~\ref{resource2} for the case $s=11\dots 11$ where the pairwise checks occur on consecutive pairs of query bits. Note that in general the pairwise checks may need to be applied between non-consecutive pairs of query bits, for example in the case $s=01 \dots 10 \dots 10$, which requires two non-consecutive checks. Thus, to implement representatives for all 1-1 and 2-1 functions using the same circuit we require $n$ query bits, $n$ ancilla bits and controllable-{\sf CNOT}'s between each query bit and all the ancilla bits except for the $n^{th}$ (as we can choose the target of the checks to be the first ancilla bit of a query pair). Then, by adding a final {\sf CNOT} between the $n^{th}$ query bit and $n^{th}$ ancilla bit we include the 1-1 functions, giving a total of $n(n-1)+1$ {\sf CNOT}'s. Each of the $2n$ bits (query and ancilla) has a corresponding qubit in the quantum resource and each {\sf CNOT} requires a {\it bridging} qubit and two edges. Thus the minimal graph resource for performing SP$_{nn}$ contains $n^2+n+1$ qubits and $2n^2-2n+2$ edges. This resource scales polynomially with $n$. Note that based on this analysis, for SP$_{22}$ we find that we require a graph resource with 7 qubits and 6 edges to implement representatives of the 1-1 and 2-1 functions. On the other hand, the 8-qubit resource shown in Fig.~\ref{resource}~(a) allows all 1-1 and 2-1 functions to be implemented. The six-qubit resource we have used to experimentally demonstrate SP$_{22}$ is a special case, with no $1-1$ functions (with the removal of 1 bridging qubit and two edges between the first query qubit and second ancilla qubit). This more compact resource still enables the performance of a quantum algorithm with a speedup over its classical counterpart~\cite{SSimon}.

\newpage

\end{widetext}


\begin{thebibliography}{99}
\bibitem{footFF} In the case of FF, based on the measurement outcomes of the qubits, the query outcomes have bit flips applied (see Tab.~\ref{tab1}) in order to retrieve the correct values.
\end{thebibliography}
\end{document}